\documentclass[letter,11pt]{article}
\pdfoutput=1 
\usepackage[utf8]{inputenc} 
\usepackage{jcappub} 
\newcommand{\be}{\begin{equation}}
\newcommand{\ee}{\end{equation}}
\newcommand{\beq}{\begin{equation}}
\newcommand{\eeq}{\end{equation}}
\newcommand{\bea}{\begin{eqnarray}}
\newcommand{\eea}{\end{eqnarray}}
\newcommand{\bear}{\begin{eqnarray}}
\newcommand{\eear}{\end{eqnarray}}

\usepackage{graphicx}  
\usepackage{dcolumn}   
\usepackage{bm,relsize}        
\usepackage{amssymb,amsmath,mathrsfs}
\usepackage{mathtools}
\usepackage{textcomp}
\usepackage{wasysym}
\usepackage{slashed}
\usepackage{multirow}
\usepackage{lipsum, color}
\usepackage[usenames,dvipsnames,svgnames]{xcolor}
\usepackage{booktabs}
\usepackage{xfrac}

\usepackage[capitalize]{cleveref}
\usepackage{subfig}

\newcommand{\bs}[1]{\boldsymbol{#1}}
\newcommand{\sinc}{\mathrm{sinc}}
\newcommand{\three}{\mathrm{\uppercase\expandafter{\romannumeral3}}}
\newcommand{\four}{\mathrm{\uppercase\expandafter{\romannumeral4}}}
\renewcommand{\d}{\mathrm{d}}

\begin{document}
\begin{flushright}
\texttt{FERMILAB-PUB-25-0140-T}
\end{flushright}

\title{Probing Primordial Power Spectrum and Non-Gaussianities With Fast Radio Bursts}
\author{Zhiyao Lu$^1$}

\author{Lian-Tao Wang$^2$}

\author{Huangyu Xiao$^{3,4}$}

\affiliation{$^1$School of Physics and State Key Laboratory of Nuclear Physics and Technology, Peking University, Beijing 100871, China}
\affiliation{$^2$ Department of Physics, Kadanoff Center for Theoretical Physics \& Enrico Fermi Institute, University of Chicago }
\affiliation{$^3$ Astrophysics Theory Department, Theory Division, Fermilab, Batavia, IL 60510, USA}
\affiliation{$^4$Kavli Institute for Cosmological Physics, University of Chicago, Chicago, IL 60637}

\emailAdd{luzhiyao@stu.pku.edu.cn}
\emailAdd{liantaow@uchicago.edu}
\emailAdd{huangyu@fnal.gov}

\abstract{ We use the precision measurements of the arrival
time differences of the same fast radio burst (FRB) source along multiple sightlines to measure the primordial power spectrum and Non-Gaussianities. The anticipated experiment requires a sightline separation of 100 AU, achieved by sending three or more radio telescopes to the outer solar system.
The Shapiro time delays, measured relatively between different telescopes, are sensitive to the gradient field of the gravitational potential between different sightlines.  Since the arrival time difference is independent of when the transient signal is emitted from the source, every measurement of the detected FRB source can be correlated. With enough FRB sources discovered, we can map the gravitational potential across the sky. We further calculate the two-point and three-point correlation function of the arrival time difference between telescopes for different FRB sources in the sky. If $10^4$ FRBs were to be detected, our results suggest that this technique can test the inflationary scale-invariant power spectrum down to $\sim 10^3\,\rm Mpc^{-1}$ and primordial Non-Gaussianities at a level of $f_{\rm NL}\sim 1$.

}

\maketitle

\section{Introduction}
The simplest single-field inflation models, with a prediction of a nearly scale-invariant adiabatic primordial scalar curvature perturbations, are largely consistent with measurements from cosmic microwave background (CMB) \cite{Planck:2018vyg} and the distribution of galaxies tracing the large scale structure (LSS) of the Universe \cite{SDSS:2003tbn,Ivanov:2019pdj}. However, the predictions of inflation models have not been extensively tested. For instance, the primordial matter power spectrum on subgalactic scales is still only weakly constrained \cite{Bechtol:2022koa}, with a few proposals that may detect the gravitational effect of dark matter subhalos \cite{Dai:2018mxx,VanTilburg:2018ykj,Dror:2019twh,Lee:2020wfn,Ramani:2020hdo,Mondino:2020rkn,Mondino:2023pnc,Dai:2020rio,Tambalo:2022wlm,Lin:2023ccz,Graham:2024hah,Dekker:2024nkb,deKruijf:2024voc}. Also, the matter power spectrum may deviate from the standard prediction in many models of dark matter \cite{Kolb:1993zz,Graham:2015rva,Eggemeier:2019khm,Eroncel:2022efc, Xiao:2021nkb, Chang:2024fol,Chung:2024ctx,Chung:2023xcv}, and early Universe dynamics \cite{Erickcek:2011us,Redmond:2018xty,Nelson:2018via,Adi:2023doe,Ralegankar:2024arh}. Moreover, while the current data suggest the primordial perturbations are consistent with a Gaussian random field \cite{Planck:2019kim}, inflation models with self-interactions or additional fields predict large primordial Non-Gaussianities that are yet to be thoroughly tested \cite{Bartolo:2004if,Bartolo:2003jx,Zaldarriaga:2003my,Meerburg:2019qqi,Maldacena:2002vr,Babich:2004gb,Chen:2006nt,Langlois:2008qf,Senatore:2009gt,Chen:2009we,Arkani-Hamed:2015bza,An:2017rwo,Peterson:2010mv,Chen:2016hrz,Chen:2016uwp,Hook:2019vcn,Kumar:2018jxz,Wang:2019gbi,Wang:2020ioa,Adshead:2021hnm,Green:2023uyz,Lodman:2023yrc}. Therefore, measuring the primordial power spectrum and Non-Gaussianities can directly probe the dynamics in the early universe and dark matter physics.

Fast Radio Burst Timing, a technique using the precision comparison of the arrival times of Fast Radio Burst (FRB) signals along multiple sightlines from the same FRB source, has been proposed to measure the Hubble constant, probe dark matter substructures, and detect gravitational waves \cite{Boone:2022pdz,Xiao:2024qay,Lu:2024yuo}. \footnote{Fast Radio Bursts, radio transients with durations at millisecond timescale, are mostly extragalactic origin. Thousands of them have been discovered \cite{CHIMEFRB:2021srp,CHIMEFRB:2023hfj}, with more than fifty known to repeat, often on timescales of tens to thousands of hours \cite{CHIMEFRB:2023myn}. While the exact nature and origin of FRBs remains unclear, fast radio bursts already provide a new window for studying cosmology \cite{McQuinn:2013tmc,Ravi:2018ose,Prochaska_2019,Heimersheim:2021meu,Li:2017mek,Zitrin:2018let,Wucknitz:2020spz,Tsai:2023tyw,Pearson:2020wxb,Dai:2017twh,Saga:2024dva,Hagstotz:2021jzu,Wu:2021jyk,Yang:2022ftm} and Beyond the Standard Model physics \cite{Munoz:2016tmg,Buckley:2020fmh,Xiao:2022hkl,Prabhu:2023cgb,Lemos:2024jbl,Gao:2023xbi,Wang:2024sdz}.} It requires either strongly lensed repeating FRBs or sending multiple radio telescopes to the outer solar system to measure the same FRB source from different sightlines. In other words, we need to achieve solar-system scale interferometry on FRBs for the proposed experiment.
The timing precision on the arrival time difference between sightlines can be subnanosecond scale through the coherent analysis of FRBs. This also requires the detectors to be located at centimeter precision. In the case of space emission, acceleration noises in the outer solar system introduce centimeter uncertainty, which could be corrected with weekly global navigation satellite system (GNSS)-like trilaterations between different detectors \cite{Boone:2022pdz,McQuinn:2024fpj}.

The advantage of FRB timing in measuring the matter power spectrum and Non-Gaussianities originates from several factors. First, the precise measurements of the arrival times make it possible to be sensitive to the gravitational potential along the line of sight. This is achieved through the coherent analysis of FRB signals from cosmological distances. FRB is the only extragalactic source that is point-like enough for this coherent analysis. Second, unlike other observations that measure the distribution of galaxies, FRB timing directly measures the total gravitational potential in the Universe, providing a clean probe of primordial physics. Previous studies on FRB timing focused on the temporal information of arrival time difference that relies on repeating FRBs \cite{Xiao:2024qay,Lu:2024yuo}, which is ideal for time-varying signals arising from small dark objects that are passing through the line-of-sight, or gravitational waves. In this work, we extend previous studies to the spatial domain and perform calculations on the spatial correlation of the timing signals from different FRB sources. Our calculation demonstrates its great sensitivity to primordial matter power spectrum and Non-Gaussianities, further strengthening the science case of this potential space mission. For the matter power spectrum, we can test the primordial perturbations down to $k\sim 10^3 \,\rm Mpc^{-1}$, which is beyond the current reach of Lyman-$\alpha$\footnote{Note that the perturbations with such wavenumber are in the nonlinear regime, which requires forward modeling of primordial density fluctuations for exact matching. In this work, we will not take into account such non-linear effects. }.
For the primordial Non-Gaussianities, we can reach a sensitivity of $f_{\rm NL}\sim 1$ for local, equilateral, and orthogonal shapes, assuming $10^4$ FRBs to be detected, and a baseline of 100 AU.

This work is organized as follows: In Sec.~\ref{sec:arrival_time_diff}, we summarize the basic techniques to compute the arrival time difference between different telescopes and show its sensitivity to the matter power spectrum. In Sec.~\ref{sec:two_point}, we use the established framework to calculate the two-point correlation function of the arrival time difference and presents its sensitivity to the matter power spectrum. In Sec.~\ref{sec:three_point}, we provide a more complicated version of this signal and provide an analytical calculation for the three-point correlation function for local type, equilateral shape, and orthogonal shape Non-Gaussianities. In Sec.~\ref{sec:conclusion}, we summarize our main result and conclude.

\section{Arrival Time Difference Between Different Detectors }\label{sec:arrival_time_diff}
\begin{figure}
    \centering
    \includegraphics[width=0.6\linewidth]{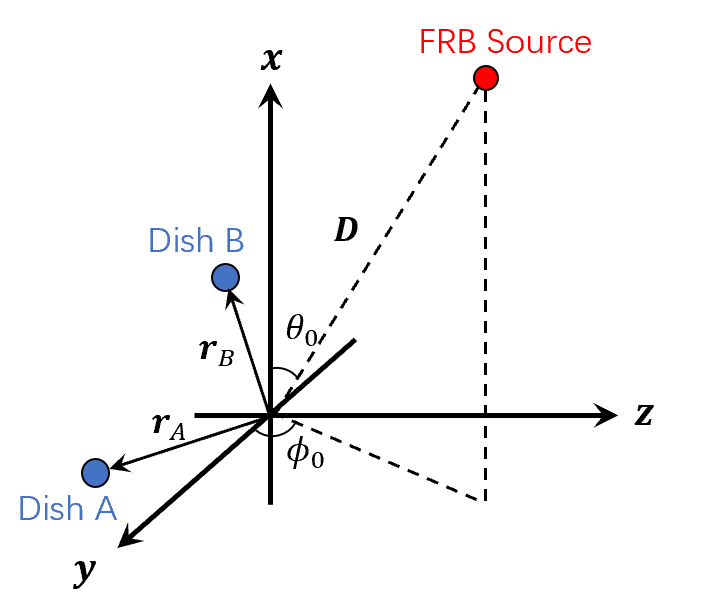}
    \caption{The geometric setup of our proposed experiment. The FRB source is at a cosmological distance and the radio dishes are anticipated to be located in the outer solar system. ($D, \theta_0,\phi_0$) defines the 3D location of the FRB source. We have chosen Earth as our reference point for convenience, which will not affect the calculation.
    }
    \label{fig:set_up}
\end{figure}

In this section, we set up the calculation of the arrival time difference between different detectors, given the gravitational potential in the Universe. Later, the correlation function of arrival time difference can be related to the power spectrum and trispectrum of the primordial curvature perturbation.
The primary observable in this experiment is the arrival time difference, which is predominantly impacted by the gradient of the gravitational potential.
Consider the following metric with a Newtonian potential $\Phi(x^\mu)$
\begin{equation}
    \d s^2=-(1+2\Phi)\d t^2+a(t)^2(1-2\Phi)\delta_{ab}\d x^a\d x^b,
\end{equation}
The Shapiro time delay along the line of light from a source at $z_i$ and observer at $z_f$ is given by
\begin{equation}\label{equ:time_pert}
    t(z_f,z_i)=-2\int_{z_i}^{z_f}\Phi(z')\d z'.
\end{equation} 
Using two detectors to observe the same FRB, the arrival time at each detector would be perturbed by a gravitational potential, leading to a difference in the arrival time at the two detectors. The geometric setup of this experiment is shown in Fig.~\ref{fig:set_up}.
Define the angular position of the FRB source as $\hat{n}$,  represented by
\begin{equation}
    \hat{n}=(\cos\theta_0,\sin\theta_0\cos\phi_0,\sin\theta_0\sin\phi_0)
\end{equation}
the definition of $\theta_0$ and $\phi_0$ are shown in Fig.~\ref{fig:set_up}. The spatial coordinates of the detectors are denoted by $\bs{r}_A$ and $\bs{r}_B$. Including spatial information of detectors in Eq.~\eqref{equ:time_pert}, we can obtain the following form
\begin{equation}
    \Delta t_{AB}=2|D\hat{n}-\bs{r}_B|\int_0^1\d s\,\Phi(sD\hat{n}-
    s\bs{r}_B)-2|D\hat{n}-\bs{r}_A|\int_0^1\d s\,\Phi(sD\hat{n}-
    s\bs{r}_A),
\end{equation}
Here, $s$ is a parameter of the light path, normalized to take its value from 0 to 1. We have omitted the geometric contribution of the arrival time difference, which can be extracted with more detectors. In practice, we should expand this expression in small $|\bs{r}_A|$ and $|\bs{r}_B|$. The first order expansion is indistinguishable from a change in the source angular position. 
Hence, the leading observable effect arises at second order, i.e., the quadrupole term, which can be achieved with more detectors. 
In other words, we are measuring the angular position of the FRB source with high precision with two detectors and then using more detectors to measure the Shapiro time delay by expanding to the quadrupole term.

\subsection*{Three Detectors}
\begin{figure}
    \centering
    \includegraphics[width=\linewidth]{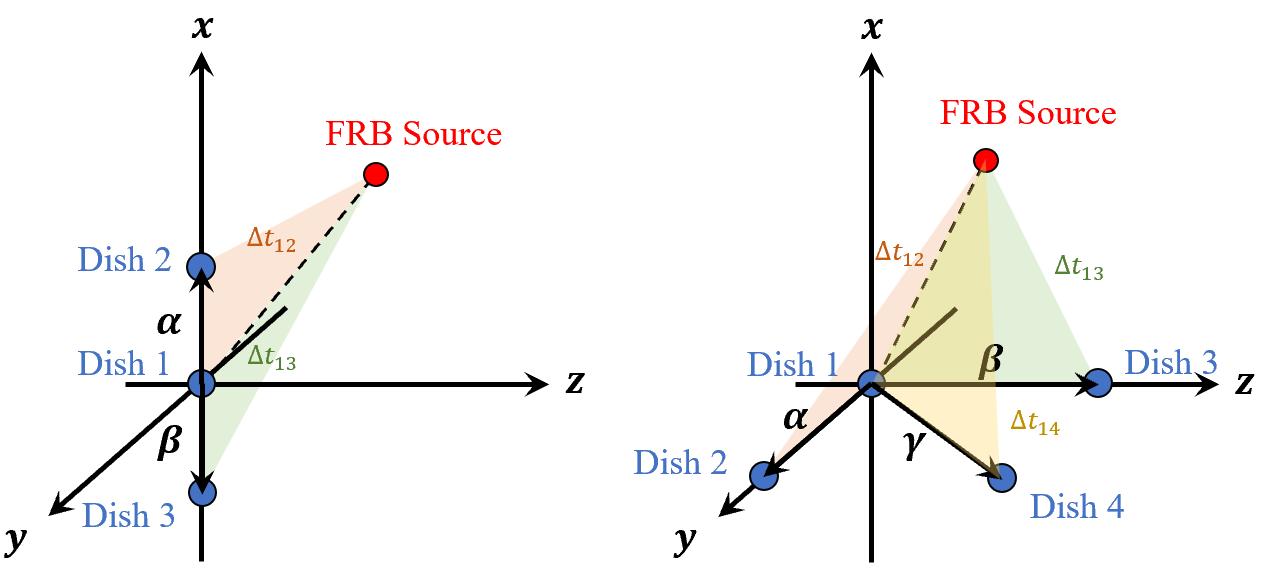}
    \caption{Schematic plot of the relative positions of the dishes. The left figure shows the three-dish case, where the three detectors should be collinear, with $\bs{\alpha}+\bs{\beta}=0$. The right figure shows the four-dish case, where the four dishes are placed on the vertices of a parallelogram, namely $\bs{\alpha}=\bs{\beta}+\bs{\gamma}$. The figure shows the special case where this parallelogram is a rectangle. }
    \label{fig:dishes}
\end{figure}
Assuming three detectors are perfectly aligned along the same line as shown in Fig.~\ref{fig:dishes}, we can obtain information about the quadrupolar term in a more tractable setup. In reality, four detectors are needed for the full information. Denote the position of the $i$th detector to be $\bs{r}_i$, we impose the following condition 
\begin{equation}
    \bs{r}_2-\bs{r}_1=\bs{\alpha},\ \bs{r}_3-\bs{r}_1=\bs{\beta},\ \bs{\beta}+\bs{\alpha}=0
\end{equation}
Consider the first order expansion of $\Delta t_{12}$ in $\bs\alpha$, 
\begin{equation}\label{eq:dt_1storder}
    \Delta t_{12}^{(1)}=2(\bs{\alpha})_a\int_0^1\d s \left[(\hat{n})_a\Phi(sD\hat{n})+sD\nabla_a\Phi(sD\hat{n})\right],
\end{equation}
where $a= 1,2,3$ labels the spatial components.
Similarly, the first order expansion of $\Delta t_{13}$ is the same as Eq.~\eqref{eq:dt_1storder} but with $\bs\alpha$ replaced by $\bs\beta$. Considering the observable $\Delta t_{12}+\Delta t_{13}$, the first order expansion is zero because $\bs\alpha+\bs\beta=0$. Therefore, to the lowest order in $\bs{\alpha}$ and $\bs{\beta}$, we have
\begin{align}\label{eq:threedish}
    &\Delta t_{12}+\Delta t_{13}\approx Q^{\three}_{ab}\nonumber\\
    &\times\int_0^1\d s\left[\frac{\delta_{ab}-n_an_b}{D}\Phi(sD\hat{n})+s\,n_a\nabla_b\Phi(sD\hat{n})+s\,n_b\nabla_a\Phi(sD\hat{n})+s^2D\nabla_a\nabla_b\Phi(sD\hat{n})\right]
\end{align}
where 
\begin{equation}
    Q^{\three}_{ab}=(\bs{\alpha})_a(\bs{\alpha})_b+(\bs{\beta})_a(\bs{\beta})_b=2\bs{\alpha}_a\bs{\alpha}_b
\end{equation}
The integrand of Eq.~\eqref{eq:threedish} can be grouped into three parts based on the order of derivatives. The derivative is taken with respect to the spatial components of $s\,\hat{n}$.
The first term is of order $\mathcal{O}\left((kD)^0\right)$, the second and third $\mathcal{O}(kD)$ and the fourth $\mathcal{O}\left((kD)^2\right)$. We are most interested in momentum ranging from $10^{-3}\,\mathrm{Mpc}^{-1}$ to $10\,\mathrm{Mpc}^{-1}$, and the distance of FRB source is roughly $D\sim 1\,\mathrm{Gpc}$, so $kD$ is always larger than 1. Therefore, the first term in the integrand makes the smallest contribution, and the fourth term makes the largest contribution.

\subsection*{Four Detectors}
Now, we will work with four detectors to compute the quadrupolar term of the arrival time difference. For convenience, we assume four coplanar detectors that form a parallelogram as shown in Fig.~\ref{fig:dishes}. Again, denote the position of the $i$th detector to be $\bs{r}_i$, we have 
\begin{equation}
    \bs{r}_2-\bs{r}_1=\bs{\alpha},\ \bs{r}_3-\bs{r}_1=\bs{\beta},\ \bs{r}_4-\bs{r}_1=\bs{\gamma},\ \bs{\alpha}=\bs{\beta}+\bs{\gamma}
\end{equation}
Consider an observable $\Delta t_{12}-\Delta t_{13}-\Delta t_{14}$. It is straightforward to see that the first-order expansion is zero. Therefore, to the lowest order in $\bs{\alpha}$, $\bs{\beta}$ and $\bs{\gamma}$, we have
\begin{align}
    &\Delta t_{12}-\Delta t_{13}-\Delta t_{14}\approx Q^{\four}_{ab}\nonumber\\
    &\times\int_0^1\d s\left[\frac{\delta_{ab}-n_an_b}{D}\Phi(s\,D\hat{n})+s\,n_a\nabla_b\Phi(s\,D\hat{n})+s\,n_b\nabla_a\Phi(sD\hat{n})+s^2D\nabla_a\nabla_b\Phi(sD\hat{n})\right]
\end{align}
where
\begin{equation}
    Q^{\four}_{ab}=(\bs{\alpha})_a(\bs{\alpha})_b-(\bs{\beta})_a(\bs{\beta})_b-(\bs{\gamma})_a(\bs{\gamma})_b
\end{equation}
One should notice that the requirement on the position of detectors can be relaxed. The difference in $|\bs{\alpha}|$ and $|\bs{\beta}|$ can be compensated by multiplying $\Delta t_{12}$ and $\Delta t_{13}$ with different coefficients. For example, if $\bs\alpha=-2\bs\beta$, we can always compute $\Delta t_{12}+2\Delta t_{13}$, so that the first order expansion is still zero. In the four detector case, we can always construct an observable $c_1\Delta t_{12}+c_2\Delta t_{13}+c_3\Delta t_{14}$ with zero first order expansion, since we can always find three real numbers $c_1$, $c_2$, and $c_3$ satisfying $c_1\bs\alpha+c_2\bs\beta+c_3\bs\gamma=0$. Essentially, four detectors contain enough information to accurately measure the 3-D location of the source, and further obtain the information on the Shapiro time delay.

\section{Two Point Correlation and Power Spectrum}\label{sec:two_point}

In this section, we will use the three dish case as an example to estimate the sensitivity of our proposed experiment, while the four dish result can be obtained using similar methods.  Before computing the correlation of arrival time differences, we first review the gravitational potential power spectrum from primordial curvature perturbations. The Fourier transform of the gravitational potential is
\begin{equation}
    \Phi(\bs{x})=\int\frac{\d^3\bs{k}}{(2\pi)^3}\tilde{\Phi}(\bs{k})e^{i\bs{k}\cdot\bs{x}}
\end{equation}
The correlation of gravitational potential in momentum space gives the power spectrum 
\begin{equation}
    \left<\tilde{\Phi}(\bs{k}_1)\tilde{\Phi}(\bs{k}_2)\right>=(2\pi)^3\delta^3(\bs{k}_1+\bs{k}_2)P_\Phi(k_1)
\end{equation}
We have assumed an isotropic Universe, so the power spectrum only depends on the magnitude of wavenumber, $k$. The power spectrum of gravitational potential can be expressed in terms of the matter overdensity power spectrum as
\begin{equation}
    P_\Phi(k)=\frac{(4\pi G\bar{\rho}_m a^2)^2 }{k^4}P_\delta(k)
\end{equation}
On superhorizon scales, we can relate the potential to the curvature through $\mathcal{R}= -3/5\,\Phi$ in matter-dominated Universe. Therefore, the matter power spectrum can be related to the curvature spectrum as
\begin{equation}
    P_\delta(k,a) = \frac{4}{25}\left(\frac{k}{H_0}\right)^4 T^2(k/k_{\rm eq}) \left(\frac{D_1(a)}{\Omega_m}\right)^2 P_{\mathcal{R}}(k).
\end{equation}
where the growth function describes the $k$-independent growth of the perturbation. The growth function is defined as
\begin{equation}
    D_1(a)=\frac{5\Omega_m}{2}\frac{H(a)}{H_0}\int_0^a\frac{da^{\prime}}{(a^{\prime}H(a^\prime)/H_0)^3},
\end{equation}
and we can approximate $H(a)/H_0\approx \sqrt{\Omega_m\, a^{-3}+1-\Omega_m}$ by ignoring radiation. The growth function is normalized in the matter-dominated Universe to be $D_1(a)=a$, and it is smaller than 1 at $a=1$ due to the cosmological constant.
The transfer function is defined as $T(k/k_{\rm eq})= \Phi(k)/\Phi(k=0)$, which characterizes the $k$-dependent growth of the gravitational potential.
$k_{\rm eq}\approx 0.073\, \Omega_m h^2\rm Mpc^{-1}$ is the inverse comoving Hubble radius at matter radiation equality. We choose the fitting formula of BBKS transfer function \cite{Bardeen:1985tr}
\begin{equation}
    T(x)=\frac{{\rm ln}(1+0.171x)}{0.171x}\left( 1+0.284 x+(1.18x)^2+(0.399x)^3+(0.490x)^4\right)^{-0.25}
\end{equation}
It is also useful to introduce the dimensionless power spectrum,
\begin{equation}
    \Delta_{\mathcal{R}}^2(k)=\frac{k^3}{2\pi^2}P_{\mathcal{R}}(k) = A_s \left(\frac{k}{k_\star}\right)^{n_s-1},
\end{equation}
where $k_\star = 0.05 \rm Mpc^{-1}$ is the pivot scale and $A_s = 2.101\times 10^{-9}$ is the amplitude of the curvature power spectrum. Now, we have a full treatment to obtain the linear matter power spectrum from the primordial curvature power spectrum.

\begin{figure}
    \centering
    \includegraphics[width=0.6\linewidth]{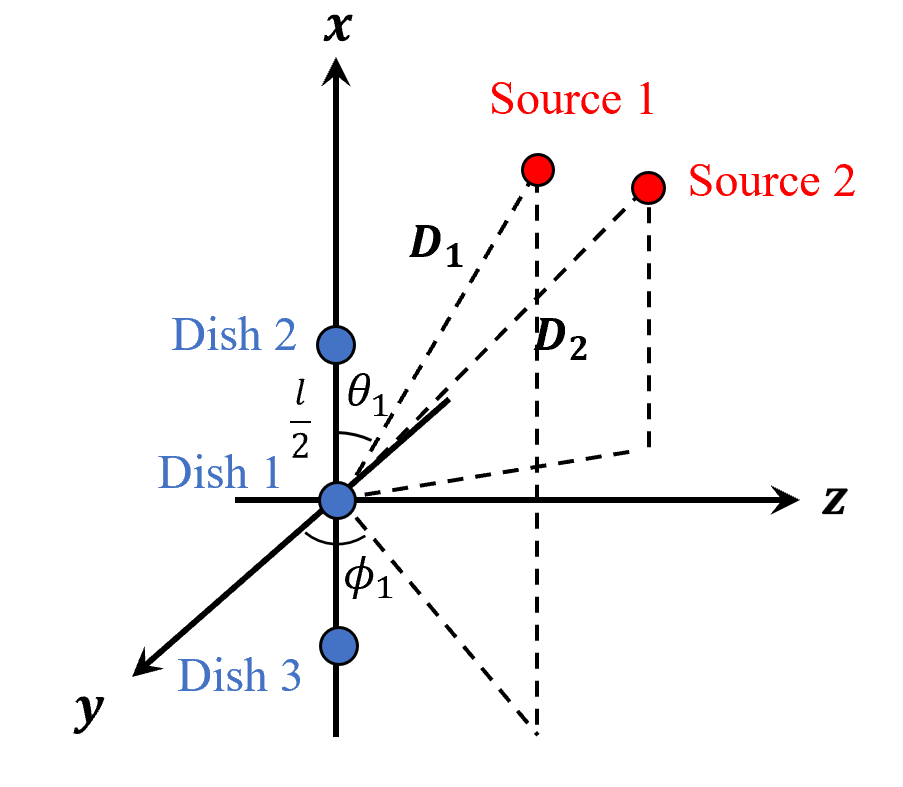}
    \caption{Measuring the two point correlation function using the three-dish setup. The distance between two adjacent dishes is $l/2$, and the distance from dish 1 to the two sources are $D_1$ and $D_2$. The angular variables $\theta_1$ and $\phi_1$ are defined as shown in the figure; $\theta_2$ and $\phi_2$ are defined in the same manner. }
    \label{fig:2s3d}
\end{figure}

Choosing dish 1 as the origin, $\bs{\alpha}$ to be $\frac{l}{2}\hat{e}_x$,  then $Q_{12}^{\three}+Q_{13}^{\three}$ has only one non-zero component, which is 
\begin{equation}(Q_{12}^{\three}+Q_{13}^{\three})_{xx}=\frac{l^2}{2}
\end{equation}
For simplicity, we will denote the lowest order expansion of $\Delta t_{12}+\Delta t_{13}$ in $l$ by $\Delta t^{(2)}$, since this quantity is proportional to $l^2$ as shown in equation \eqref{eq:threedish}. The explicit expression of $\Delta t^{(2)}$ is
\begin{align}\label{eq:delta_t_2}
  \Delta t_{12}+\Delta t_{13} \xrightarrow{\rm lowest\, order} \Delta t^{(2)}= \frac{l^2}{2}\int_0^1\d s\left[\frac{1-n_x^2}{D}\Phi(sD\hat{n})+2s\ n_x\nabla_x\Phi(sD\hat{n})+s^2D\nabla_x^2\Phi(sD\hat{n})\right]
\end{align}
It is straightforward to compute the two-point correlation function of the arrival time difference, which gives 
\begin{align}\label{eq:corr_3dish}
    &\left<\Delta t^{(2)}(D_1,\theta_1,\phi_1)\Delta t^{(2)}(D_2,\theta_2,\phi_2)\right>=\frac{l^4}{4D_1D_2}\int\frac{\d^3\bs{k}}{(2\pi)^3}P_\Phi(k)e^{i(k_1D_1-k_2D_2)/2}\nonumber\\
    &\left[\sinc\left(\frac{k_1D_1}{2}\right)\sin^2\theta_1-k_xD_1f\left(\frac{k_1D_1}{2}\right)\cos\theta_1-\frac{k_x^2D_1^2}{2}g\left(\frac{k_1D_1}{2}\right)\right]\nonumber\\
    &\left[\sinc\left(\frac{k_2D_2}{2}\right)\sin^2\theta_2-k_xD_2f^*\left(\frac{k_2D_2}{2}\right)\cos\theta_2-\frac{k_x^2D_2^2}{2}g^*\left(\frac{k_2D_2}{2}\right)\right]
\end{align}
where $k_1=\bs k \cdot\hat n_1$, $k_2=\bs k \cdot\hat n_2$, and $\sinc(x)=\sin(x)/x$. This expression is derived by expanding all potentials in its Fourier modes, substituting the power spectrum for the correlation function between different modes, and carrying out the $s$ integral. We have defined two functions
\begin{equation}
    f(x)=\frac{\sinc(x)-e^{ix}}{x}
\end{equation}
\begin{equation}
    g(x)=\frac{-\sinc(x)+e^{ix}-ixe^{ix}}{x^2}
\end{equation}
Notice that although these two functions are superficially divergent at $x=0$, one would find they are finite by studying the behavior with a small $x$. The integration over momentum $\bs k$ should have an infrared cutoff at Hubble, which is roughly $10^{-3}\mathrm{Mpc}^{-1}$. 

The integral in Eq.~\eqref{eq:corr_3dish} is very expensive to evaluate, so we use some approximations to simplify the calculation. Since $D$ is around 1Gpc, relevant wavenumber satisfies $|\bs k|D\gtrsim 1$. Note that the three terms in one square bracket have different powers of $k_x D$. Thus, the greatest contribution comes from the $k_x^2D^2$ term, and the first two terms in each square bracket can be neglected. 

\begin{figure}[h]
    \centering
    \includegraphics[width=0.9\linewidth]{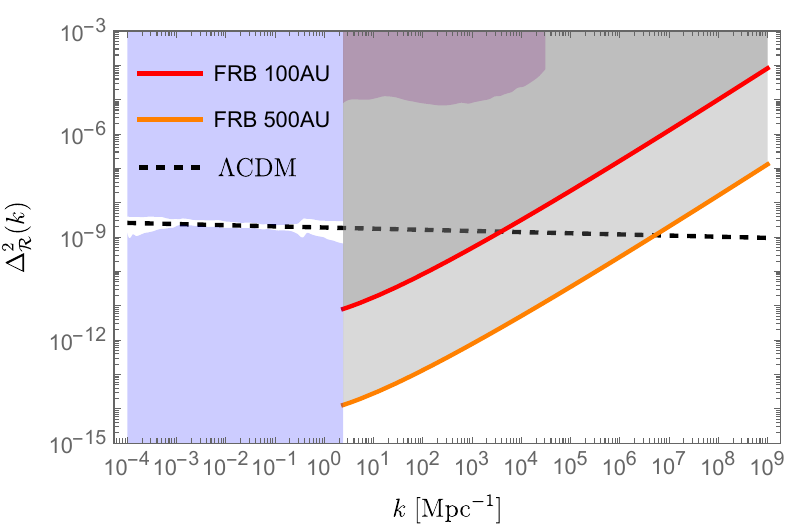}
    \caption{Our detection limit compared to the Primordial power spectrum. The red curve is from FRB timing measurement with a 100 AU baseline, and the orange curve a 500 AU baseline. The constraint on a longer length scale is from CMB and Lyman-$\alpha$, which is shown in blue. The purple shaded region is from CMB distortion, and the black dashed line is from the prediction of simple slow roll inflation with an approximately constant tilt. Some inflation models feature an enhanced power spectrum on small scales, which will be even more detectable. While we plot the linear power spectrum expected from inflation, FRB timing directly measures the nonlinear power spectrum in the late Universe, which requires forward modeling to exactly match to the sensitivity on the linear power spectrum.
   }
    \label{fig:powerspectrum}
\end{figure}

Also note that the function $g(x)$ has a peak at $x=0$, and is then suppressed by $1/x$ as $x$ becomes large. Therefore, the dominant contribution to this integral should come from the regime where $k_1D_1\sim 0$ and $k_2 D_2\sim 0$, corresponding to $\bs k\perp \hat n_1$ and $\bs k\perp \hat n_2$. If $\hat n_1$ and $\hat n_2$ are collinear, then the phase space corresponding to the greatest contribution in the integral is maximized. Therefore, we expect the strongest correlation to come from the power of a single source, which is simply
\begin{equation}\label{eq:beforelimber_2point}
    \left<\Delta t^{(2)}(D,\theta)\Delta t^{(2)}(D,\theta)\right>=\frac{l^4D^2}{16}\int\frac{\d^3\bs{k}}{(2\pi)^3}P_\Phi(k)k_x^4\left|g\left(\frac{k_nD}{2}\right)\right|^2
\end{equation}
Since the primary contribution to the integral comes from $k_n\approx 0$, we may approximate $|g(k_nD/2)|^2$ by a delta function. 
\begin{equation}
    |g(ax)|^2\approx \frac{4\pi}{5a}\delta(x)\mbox{ for large }a
\end{equation}
Then, we can obtain the analytical form of the autocorrelation function 
\begin{equation}\label{eq:limber_2point}
    \left<\Delta t^{(2)}(D,\theta)\Delta t^{(2)}(D,\theta)\right>=\frac{3l^4D}{160}\int\frac{\d k}{2\pi}k^5P_\Phi(k)\sin^4\theta.
\end{equation}
While the observation time is not long enough to see temporal changes of contributions from long-wavelength modes, the autocorrelation function predicts the expected power of $\Delta t^{(2)}(D,\theta)$. 
Using the inflationary power spectrum, and taking benchmark values $D=3\,\mathrm{Gpc}$, $l=100\,\mathrm{AU}$, we have
\begin{equation}
    \left<\Delta t^{(2)}(D,\theta)\Delta t^{(2)}(D,\theta)\right>=3.47\times 10^{-21}\mathrm{s}^2\times \sin^4\theta
\end{equation}
Assuming the signal falls primarily in one logarithmic bin of wave number $k$, averaging the signal over $\theta$, we can estimate the sensitivity to the power spectrum of primordial curvature perturbation on that scale to a precision
\begin{equation}
    \delta P_\mathcal{R}(k)\approx \frac{200\pi \delta t_m^2}{Nl^4Dk^6}\times \left(\frac{4}{25}T(k/k_{\rm eq})^{2}D_1(a=1)^{2}\right)^{-1}
\end{equation}
where we have assumed a uniform distribution of FRB sources in the sky. $N$ is the total number of FRB sources, and $\delta t_m$ is the FRB timing precision. $T(k)$ is the transfer function, and $D_1(a)$ is the growth function. We include these two functions to convert the power spectrum today to the primordial power spectrum. In Fig.~\ref{fig:powerspectrum}, we plot the sensitivity of our probe in comparison to the primordial power spectrum, showing that the FRB timing technique can greatly expand the current reach on the matter power spectrum. Note that our probe will be sensitive to extremely small scales where the primordial perturbations are nonlinear. Our observable, which measures the gravitational potential in the late Universe, is directly determined by the nonlinear power spectrum.
Therefore, more accurate forward modeling is required to match the primordial power in the nonlinear regime, while we only present the linear power in this work for demonstration.  

The two-point correlation function of the arrival time difference between two separate sources is also worth studying. Although the signal is not as large as the power of time delay difference for a single source, the statistics are enhanced by a factor equal to the total number of FRB pairs. However, Eq.~\eqref{eq:corr_3dish} is a three-dimensional integral with an intimidating integrand, which is difficult for numerical computation, and also impossible to analyze how the signal changes with different geometries. Instead of tackling this integral by brute force, we will introduce a trick to compute the correlation function of separate sources. 
Start with the time difference
\begin{equation}
    \Delta t^{(2)}(\hat{n})=\frac{l^2}{2D^2}\int_0^D\d r\int\frac{\d^3\bs{k}}{(2\pi)^3}\tilde{\Phi}(\bs{k})\left(1-n_x^2+2irn_xk_x-r^2k_x^2\right)e^{ir\bs{k}\cdot\hat{n}}
\end{equation}
The time difference is a function of the unit vector $\hat{n}$. However, it can also be treated as a function of three free variables $n_x$, $n_y$, and $n_z$ if we ignore the normalization condition $\hat{n}^2=1$. Specifically, we still use $\hat{n}$ as a unit vector but introduce an auxiliary vector $\bs{\eta}=\eta \hat{n}$. The three components of $\eta$ are free variables, and we rewrite the time difference as
\begin{equation}
    \Delta t^{(2)}(\bs{\eta})=\frac{l^2}{2D^2}\int_0^D\d r\int\frac{\d^3\bs{k}}{(2\pi)^3}\tilde{\Phi}(\bs{k})\left(1-\eta_x^2+2ir\eta_xk_x-r^2k_x^2\right)e^{ir\bs{k}\cdot\bs{\eta}}
\end{equation}
Observe that the terms $rk_x$ and $r^2k_x^2$ can be replaced by the derivative of $e^{ir\bs{k}\cdot\bs{\eta}}$ over $\eta_x$, so
\begin{equation}
    \Delta t^{(2)}(\bs{\eta})=\frac{l^2}{2D^2}\left(1-\eta_x^2+2\eta_x\frac{\partial}{\partial \eta_x}+\frac{\partial^2}{\partial \eta_x^2}\right)\int_0^D\d r\int\frac{\d^3\bs{k}}{(2\pi)^3}\tilde{\Phi}(\bs{k})e^{ir\bs{k}\cdot\bs{\eta}}
\end{equation}
For notational simplicity, we introduce differential operators 
\begin{equation}\label{eq:Oj}
    O_j=\left(1-\eta_{jx}^2+2\eta_{jx}\frac{\partial}{\partial \eta_{jx}}+\frac{\partial^2}{\partial \eta_{jx}^2}\right), j=1,2
\end{equation}
and the two point function can be written as
\begin{align}
    \left<\Delta t^{(2)}(\bs{\eta}_1)\Delta t^{(2)}(\bs{\eta}_2)\right>=\frac{l^4}{4D_1^2D_2^2}O_1O_2\int_0^{D_1}\d r_1\int_0^{D_2}\d r_2\int\frac{\d^3\bs{k}}{(2\pi)^3}P_\Phi(k)e^{i\bs{k}\cdot(r_1\bs{\eta}_1-r_2\bs{\eta}_2)}
\end{align}

Notice that $[O_1,O_2]=0$. Since $D$ is large, we can extend the upper limit of the $r_1$, $r_2$ integration to infinity. We integrate out $r_1$ and $r_2$, the result is
\begin{align}
    \left<\Delta t^{(2)}(\bs{\eta}_1)\Delta t^{(2)}(\bs{\eta}_2)\right>=\frac{l^4}{4D_1^2D_2^2}O_1O_2\int\frac{\d^3\bs{k}}{(2\pi)^3}P_\Phi(k)\left[\pi\delta(\bs{k}\cdot\bs{\eta}_1)+\frac{1}{i\bs{k}\cdot\bs{\eta}_1}\right]\nonumber\\
    \times \left[\pi\delta(\bs{k}\cdot\bs{\eta}_2)+\frac{1}{i\bs{k}\cdot\bs{\eta}_2}\right]
\end{align}
There are four terms after expanding the square brackets. Notice that the delta functions are even in $\bs{k}$, so the imaginary components vanish. Introduce three vectors
\begin{equation}
    \bs{\gamma}_1=\frac{\bs{\eta}_1}{(\bs{\eta}_1\times \bs{\eta}_2)^2},\ \bs{\gamma}_2=\frac{\bs{\eta}_2}{(\bs{\eta}_1\times \bs{\eta}_2)^2},\ \bs{\gamma}_3=\frac{\bs{\eta}_1\times \bs{\eta}_2}{(\bs{\eta}_1\times \bs{\eta}_2)^2}
\end{equation}
Perform a change of variables, 
\begin{equation}
    \rho_1=\bs{k}\cdot\bs{\gamma}_1,\ \rho_2=\bs{k}\cdot\bs{\gamma}_2,\ 
    \rho_3=\bs{k}\cdot\bs{\gamma}_3
\end{equation}
The change of measure can be computed from the Jacobian,
\begin{equation}
    \d\rho_1\d\rho_2\d\rho_3=(\bs{\eta}_1\times \bs{\eta}_2)^{-4}\d^3\bs{k}
\end{equation}
To obtain the inverse transformation, introduce three vectors
\begin{equation}
    \bs{g}_1=\bs{\eta}_2\times(\bs{\eta}_1\times \bs{\eta}_2),\ \bs{g}_2=(\bs{\eta}_1\times \bs{\eta}_2)\times \bs{\eta}_1,\ \bs{g}_3=\bs{\eta}_1\times\bs{\eta}_2
\end{equation}
We can easily obtain the following conditions
\begin{equation}
\bs{g}_i\cdot\bs{\gamma}_j=\delta_{ij},\quad   \bs{k}=\rho_1\bs{g}_1+\rho_2\bs{g}_2+\rho_3\bs{g}_3
\end{equation}
Now, the two-point correlation function can be rewritten as
\begin{equation}
    \left<\Delta t^{(2)}(\bs{\eta}_1)\Delta t^{(2)}(\bs{\eta}_2)\right>=\frac{l^4}{4D_1^2D_2^2}O_1O_2\int\frac{\d^3\bs{\rho}}{(2\pi)^3}P_\Phi(|\rho_1\bs{g}_1+\rho_2\bs{g}_2+\rho_3\bs{g}_3|)\left[\pi^2\delta(\rho_1)\delta(\rho_2)-\frac{1}{\rho_1\rho_2}\right]
\end{equation}
Though $\frac{1}{\rho_1\rho_2}$ is an odd function of both $\rho_1$ and $\rho_2$, $P_\Phi(k)$ is not, because The norm of $\bs{k}$ is
\begin{equation}
    k=|\bs{\eta}_1\times\bs{\eta}_2|\sqrt{\eta_2^2\rho_1^2+\eta_1^2\rho_2^2+\rho_3^2-2\rho_1\rho_2\bs{\eta}_1\cdot\bs{\eta}_2}
\end{equation}
Swapping the sign of $\rho_1$ would cause a change in the $\rho_1 \rho_2$ term. This means effectively we are integrating the difference of $P_\Phi$ at two different points, and this difference would be zero at $\bs{k}=0$. Therefore, the $1/(\rho_1 \rho_2)$ term introduces additional suppression at small $k$, but the function $P_\Phi(k)$ is monotonically decreasing with $k$, so this contribution should be much smaller than the delta function term. Note that when $\bs{\eta}_1\perp \bs{\eta}_2$, the $1/(\rho_1\rho_2)$ term would be exactly zero, and its contribution would gradually grow larger as the direction of $\bs{\eta}_1$ aligns with $\bs{\eta}_2$.

Keeping the first term that contains delta functions only, we have
\begin{equation}
    \left<\Delta t^{(2)}(\bs{\eta}_1)\Delta t^{(2)}(\bs{\eta}_2)\right>=\frac{l^4}{8D_1^2D_2^2}O_1O_2\frac{1}{|\bs{\eta}_1\times\bs{\eta}_2|}\int_{k>0}\frac{\d k}{2\pi}P_\Phi(k)
\end{equation}
The integral is simple to compute numerically, and the term $O_1O_2\frac{1}{|\bs{\eta}_1\times\bs{\eta}_2|}$ can be calculated analytically. Therefore, this formula can be used to numerically compute the two point correlation signal from two separate sources.

\section{Three Point Correlation and Non-Gaussianity}\label{sec:three_point}
In this section, we will use the same strategy in the previous section to compute the expected signal, namely the three-point correlation function of the arrival time difference, from primordial Non-Gaussianities.
The amplitude of the three-point correlation of the gravitational potential can be parameterized by the bispectrum $B_\Phi(p,q,l)$
\begin{equation}
    \left<\tilde{\Phi}(\bs{p})\tilde{\Phi}(\bs{q})\tilde{\Phi}(\bs{l})\right>=(2\pi)^3f_{\mathrm{NL}}B_\Phi(p,q,l)\delta^3(\bs{p}+\bs{q}+\bs{l})
\end{equation}
There are several different models of $B_\Phi(p,q,l)$, the local type is 
\begin{equation}
    B^{\mathrm{loc}}_\Phi(p,q,l)=2A^2\left[\frac{1}{p^{4-n_s}q^{4-n_s}}+\frac{1}{p^{4-n_s}l^{4-n_s}}+\frac{1}{q^{4-n_s}l^{4-n_s}}\right]\times T(p)T(q)T(l)D_1(a=1)^3
\end{equation}
where $T(p)$ is the transfer function, $A$ is the amplitude of the primordial power spectrum. The equilateral shape reads
\begin{align}
     B^{\mathrm{equil}}_\Phi(p,q,l)=&6A^2\left[-\frac{2}{(pql)^{2(4-n_s)/3}}-\left(\frac{1}{p^{4-n_s}q^{4-n_s}}+\mbox{cyclic}\right)\right.\nonumber\\
     &\left.+\left(\frac{1}{p^{(4-n_s)/3}q^{2(4-n_s)/3}l^{4-n_s}}+\mbox{5 perms}\right)\right]\times T(p)T(q)T(l)D_1(a=1)^3
\end{align}
and the orthogonal shape reads
\begin{align}
     B^{\mathrm{ortho}}_\Phi(p,q,l)=&6A^2\left[-\frac{8}{(pql)^{2(4-n_s)/3}}-\left(\frac{3}{p^{4-n_s}q^{4-n_s}}+\mbox{cyclic}\right)\right.\nonumber\\
     &\left.+\left(\frac{3}{p^{(4-n_s)/3}q^{2(4-n_s)/3}l^{4-n_s}}+\mbox{5 perms}\right)\right]\times T(p)T(q)T(l)D_1(a=1)^3
\end{align}
Given the form of arrival time difference as a function of the gravitational potential as in Eq.~\ref{eq:delta_t_2}, we can easily compute the three point correlation of time difference
\begin{align}\label{eq:bispec_3dish}
    &\left<\Delta t^{(2)}(D_1,\theta_1,\phi_1)\Delta t^{(2)}(D_2,\theta_2,\phi_2)\Delta t^{(2)}(D_3,\theta_3,\phi_3)\right>=\nonumber\\
    &f_{\mathrm{NL}}\frac{l^6}{8D_1D_2D_3}\int\frac{\d^3\bs{p}}{(2\pi)^3}\frac{\d^3\bs{q}}{(2\pi)^3}\frac{\d^3\bs{l}}{(2\pi)^3}(2\pi)^3B_\Phi(p,q,l)\delta^3(\bs{p}+\bs{q}+\bs{l})e^{i(p_1D_1+q_2D_2+l_3D_3)/2}\nonumber\\
    &\left[\sinc\left(\frac{p_1D_1}{2}\right)\sin^2\theta_1-p_xD_1f\left(\frac{p_1D_1}{2}\right)\cos\theta_1-\frac{p_x^2D_1^2}{2}g\left(\frac{p_1D_1}{2}\right)\right]\nonumber\\
    &\left[\sinc\left(\frac{q_2D_2}{2}\right)\sin^2\theta_2-q_xD_2f\left(\frac{q_2D_2}{2}\right)\cos\theta_2-\frac{q_x^2D_2^2}{2}g\left(\frac{q_2D_2}{2}\right)\right]\nonumber\\
    &\left[\sinc\left(\frac{l_3D_3}{2}\right)\sin^2\theta_3-l_xD_3f\left(\frac{l_3D_3}{2}\right)\cos\theta_3-\frac{l_x^2D_3^2}{2}g\left(\frac{l_3D_3}{2}\right)\right]
\end{align}
where $p_1=\bs p\cdot \hat n_1$, $q_2=\bs q\cdot \hat n_2$ and $l_3=\bs l\cdot \hat n_3$. The bispectrum $B_\Phi$ can be taken to be the expression of any of the three models. Following similar reasoning, we keep only the $g(x)$ term in each square bracket and focus on the single source signal
\begin{align}
    &\left<\Delta t^{(2)}(D,\hat n)\Delta t^{(2)}(D,\hat n)\Delta t^{(2)}(D,\hat n)\right>=\nonumber\\
    &-f_{\mathrm{NL}}\frac{l^6D^3}{64}\int\frac{\d^3\bs{p}}{(2\pi)^3}\frac{\d^3\bs{q}}{(2\pi)^3}B_\Phi(p,q,l)p_x^2q_x^2l_x^2g\left(\frac{p_nD}{2}\right)g\left(\frac{q_nD}{2}\right)g\left(-\frac{(p_n+q_n)D}{2}\right)
\end{align}
where $p_n=\bs p\cdot \hat n$, and similarly for $q_n$ and $l_n$. Again, we approximate the product of three functions by delta functions,
\begin{equation}\label{eq:approx_three_g}
    g(ax)g(ay)g(-a(x+y))\approx \frac{8\pi^2}{7a^2}\delta(x)\delta(y)\mbox{ for large }a
\end{equation}
Set $\hat{n}$ in the direction of $x$, the three point correlation can be simplified to be
\begin{align}
    \left<\left(\Delta t^{(2)}(D,\theta)\right)^3\right>=\frac{-f_{\mathrm{NL}}l^6 D}{896(2\pi)^3}\int\d p\d q\d\phi \ B_\Phi\left(p,q,\sqrt{p^2+q^2+2pq\cos\phi}\right)\nonumber\\
    \times p^3q^3[pq(4\cos^3\phi+6\cos\phi)+(p^2+q^2)(1+4\cos^2\phi)]\sin^6\theta
\end{align}
Numerically, the three-point correlation expected from a single source is
\begin{equation}
    \left<\left(\Delta t^{(2)}(D,\theta)\right)^3\right>=-3.55\times 10^{-38}\mathrm{s}^3\times f_{\mathrm{NL}}^{\mathrm{loc}}\left(
    \frac{l}{100\,\mathrm{AU}}\right)^6\left(
    \frac{D}{1\,\mathrm{Gpc}}\right)\times \sin^6\theta.
\end{equation}
Assuming an equilateral shape, then it becomes
\begin{equation}
    \left<\left(\Delta t^{(2)}(D,\theta)\right)^3\right>=- 1.28\times 10^{-39}\mathrm{s}^3\times f_{\mathrm{NL}}^{\mathrm{equil}}\left(
    \frac{l}{100\,\mathrm{AU}}\right)^6\left(
    \frac{D}{1\,\mathrm{Gpc}}\right)\times  \sin^6\theta.
\end{equation}
We can also perform a similar calculation for the orthogonal shape, which gives
\begin{equation}
    \left<\left(\Delta t^{(2)}(D,\theta)\right)^3\right>= 1.07\times 10^{-38}\mathrm{s}^3\times f_{\mathrm{NL}}^{\mathrm{ortho}}\left(
    \frac{l}{100\,\mathrm{AU}}\right)^6\left(
    \frac{D}{1\,\mathrm{Gpc}}\right)\times \sin^6\theta.
\end{equation}
We can sum up the signal from different sources to enhance the statistics. With $N$ sources evenly distributed in $4\pi$ solid angle, the $\sin^6\theta$ term should be replaced by its average $\frac{16}{35}$, and the signal is multiplied by $N$. We compare this signal with the detection sensitivity $\delta t_m^3$, and therefore obtain a detection limit on $f_{\mathrm{NL}}$. If we take $D=3\,\mathrm{Gpc}$, $l=500\,\mathrm{AU}$, $N=10^4$, we can detect $f_{\mathrm{NL}}$ with a precision of $10^{0}$ or $10^{-1}$, depending on the model we use.

\begin{subequations}
    \begin{align}
        &\left|f_{\mathrm{NL}}^{\mathrm{loc}}\right|=1.31\times 10^{-1}\times \left(\frac{10^4}{N}\right)\left(\frac{500\mathrm{AU}}{l}\right)^6\left(\frac{3\,\mathrm{Gpc}}{D}\right),\\
        &\left|f_{\mathrm{NL}}^{\mathrm{equil}}\right|=3.66\times \left(\frac{10^4}{N}\right)\left(\frac{500\mathrm{AU}}{l}\right)^6\left(\frac{3\,\mathrm{Gpc}}{D}\right),\\
        &\left|f_{\mathrm{NL}}^{\mathrm{ortho}}\right|=4.35\times 10^{-1}\times \left(\frac{10^4}{N}\right)\left(\frac{500\mathrm{AU}}{l}\right)^6\left(\frac{3\,\mathrm{Gpc}}{D}\right)
    \end{align}
\end{subequations}

Similar to two point correlation, the signal from three separate sources is also interesting, and there is an approximation formula to compute it. Define differential operators $O_j$ with $j=1,2,3$ as in Eq.~\eqref{eq:Oj}, the three point correlation can be written as
\begin{align}
    \left<\Delta t^{(2)}(\bs{\eta}_1)\Delta t^{(2)}(\bs{\eta}_2)\Delta t^{(2)}(\bs{\eta}_3)\right>=\frac{l^6}{8D_1^2D_2^2D_3^2}O_1O_2O_3\int_0^{D_1}\d r_1\int_0^{D_2}\d r_2\int_0^{D_3}\d r_3\nonumber\\
    \int\frac{\d^3\bs{k}}{(2\pi)^3}\frac{\d^3\bs{p}}{(2\pi)^3}B_\Phi(k,p,|\bs{k}+\bs{p}|)\exp[ir_1\bs{k}\cdot\bs{\eta}_1+ir_2\bs{p}\cdot\bs{\eta}_2-ir_3(\bs{k}+\bs{p})\cdot\bs{\eta}_3]
\end{align}
Introduce three vectors,
\begin{equation}
    \bs{\gamma}_1=\frac{\bs{\eta}_1}{v},\ \bs{\gamma}_2=\frac{\bs{\eta}_2}{v},\ \bs{\gamma}_3=\frac{\bs{\eta}_3}{v},\ v=\bs{\eta}_1\cdot(\bs{\eta}_2\times\bs{\eta}_3)
\end{equation}
and correspondingly
\begin{equation}
    \bs{g}_1=\bs{\eta}_2\times\bs{\eta}_3,\ \bs{g}_2=\bs{\eta}_3\times\bs{\eta}_1,\ \bs{g}_3=\bs{\eta}_1\times\bs{\eta}_2
\end{equation}
It is easy to verify
\begin{equation}
    \bs{\gamma}_i\cdot\bs{g}_j=\delta_{ij}
\end{equation}
Perform a change of variables, 
\begin{equation}
    \rho_1=\bs{k}\cdot\bs{\gamma}_1,\ \rho_2=\bs{k}\cdot\bs{\gamma}_2,\ \rho_3=\bs{k}\cdot\bs{\gamma}_3
\end{equation}
\begin{equation}
    \sigma_1=\bs{p}\cdot\bs{\gamma}_1,\ \sigma_2=\bs{p}\cdot\bs{\gamma}_2,\ \sigma_3=\bs{p}\cdot\bs{\gamma}_3
\end{equation}
The inverse is
\begin{equation}
    \bs{k}=\rho_1\bs{g}_1+\rho_2\bs{g}_2+\rho_3\bs{g}_3,\ \bs{p}=\sigma_1\bs{g}_1+\sigma_2\bs{g}_2+\sigma_3\bs{g}_3
\end{equation}
Similar to what we have done with the two point correlation signal, we integrate out $r_1$, $r_2$, and $r_3$ and keep only the delta function term, the three point correlation signal can be simplified into
\begin{align}
    &\left<\Delta t^{(2)}(\bs{\eta}_1)\Delta t^{(2)}(\bs{\eta}_2)\Delta t^{(2)}(\bs{\eta}_3)\right>=\frac{l^6}{64D_1^2D_2^2D_3^2}O_1O_2O_3\nonumber\\
    &\int\frac{\d\sigma_1\d\rho_2\d\rho_3}{(2\pi)^3}vB_\Phi(|\rho_2\bs{g}_2+\rho_3\bs{g}_3|,|\sigma_1\bs{g}_1-\rho_3\bs{g}_3|,|\sigma_1\bs{g}_1+\rho_2\bs{g}_2|)
\end{align}
We can begin by computing the differential operators and then performing the integration.  The resulting integrand can always be expressed polynomials in $\eta$ and derivatives of $B_\Phi$. This approach reduces the original six-dimensional integral to a three-dimensional one, enabling faster numerical computation. 
Computing the distribution of the signal for all possible positions of the three sources remains challenging. However, based on the calculation of the signal for various different values of \(\hat{n}_1\), \(\hat{n}_2\), and \(\hat{n}_3\), we estimate that the average absolute value is approximately four orders of magnitude smaller than the result obtained from auto-correlation.

Though the signal is weaker, the total number of samples of 3-point correlators is now approximately $N^3/6$ instead of $N$, which is a great enhancement with $N\sim 10^4$. With this method, we can achieve an interesting sensitivity with a smaller baseline. Assuming $D=3\,\mathrm{Gpc}$, $l=100\,\mathrm{AU}$ and $N=10^4$, we estimate our sensitivity to be 
\begin{subequations}
    \begin{align}
        &\left|f_{\mathrm{NL}}^{\mathrm{loc}}\right|\approx 1\times\left(\frac{10^4}{N}\right)^3\left(\frac{100\,\mathrm{AU}}{l}\right)^6\left(\frac{3\,\mathrm{Gpc}}{D}\right),\\
        &\left|f_{\mathrm{NL}}^{\mathrm{equil}}\right|\approx 10\times\left(\frac{10^4}{N}\right)^3\left(\frac{100\,\mathrm{AU}}{l}\right)^6\left(\frac{3\,\mathrm{Gpc}}{D}\right),\\
        &\left|f_{\mathrm{NL}}^{\mathrm{ortho}}\right|\approx 1\times\left(\frac{10^4}{N}\right)^3\left(\frac{100\,\mathrm{AU}}{l}\right)^6\left(\frac{3\,\mathrm{Gpc}}{D}\right).
    \end{align}
\end{subequations}
Note that a 10-meter dish can detect up to $10^3$ FRBs per year \cite{Boone:2022pdz}. Hence, an observation time of 10 years can detect $10^4$ FRBs that provide the desired sensitivity. Notably, the sensitivity on $f_{\rm NL}$ is proportional to $l^{-6}$. Therefore, a longer baseline will greatly enhance our sensitivity, which further motivates a space mission on probing the primordial Non-Gaussianities.

\section{Conclusion}\label{sec:conclusion}
In this work, we propose using Fast Radio Burst (FRB) timing to probe the primordial power spectrum and Non-Gaussianities. Our anticipated experiment requires a solar-system scale interferometry on Fast Radio Bursts. Using high-precision measurements of the FRB arrival time difference between different radio telescopes separated by solar-system scale, we demonstrate that FRB timing is a great probe of gravitational potential in the Universe with high sensitivity. Our calculations indicate that with a sufficiently large number of observed FRBs and an appropriate spatial separation of detectors ($\sim$100 AU), this technique can achieve great precision in constraining both the scale-invariant primordial power spectrum and primordial Non-Gaussianities. Our method can test the primordial matter power spectrum predicted by inflation down to scales of $\sim 10^3\, \rm Mpc^{-1}$ and detect primordial Non-Gaussianities at a sensitivity level of $f_{\rm NL}\sim 1$ with a 100 AU baseline and $10^4$ observed FRBs. In particular, our probe of Non-Gaussianities directly measures the three-point correlation function without introducing galaxy bias, which complements other detections using galaxy surveys.

Our calculation further strengthens the science case of FRB timing, which has been proposed to measure the Hubble constant, dark matter substructures, and gravitational waves. By directly measuring the gravitational potential, FRB timing is a complementary approach to traditional cosmological probes of the primordial power spectrum and Non-Gaussianities like the CMB and large-scale structure surveys, providing a novel window into the physics of the early universe and dark matter.

\section*{Acknowledgement}
The work of L.T.W. is supported by DOE grant
DE-SC-0013642. HX is supported by Fermi Forward Discovery Group, LLC under Contract No. 89243024CSC000002 with the U.S. Department of Energy, Office of Science, Office of High Energy Physics.
\bibliographystyle{jhep}
\bibliography{ref}

\end{document}